\documentclass[12pt,preprint]{aastex}

\shorttitle{Progressive Star Formation in NGC 602}
\shortauthors{Carlson et al.}

\begin{document}

\title{Progressive star formation in the young SMC cluster NGC 602}

\author{Lynn Redding Carlson}
\affil{Department of Physics and Astronomy, Johns Hopkins University, Baltimore, MD}
\email{carlson@stsci.edu}
\author{E. Sabbi} 
\affil{Space Telescope Science Institute, Baltimore, MD}
\email{sabbi@stsci.edu}
\author{M. Sirianni\altaffilmark{1}}
\affil{Space Telescope Science Institute, Baltimore, MD}   
\email{sirianni@stsci.edu}
\author{J. L. Hora} 
\affil{Harvard-Smithsonian Center for Astrophysics, Cambridge, MA}
\email{jhora@cfa.harvard.edu}
\author{A. Nota\altaffilmark{1}}
\affil{Space Telescope Science Institute, Baltimore, MD}   
\email{nota@stsci.edu}
\author{M. Meixner}
\affil{Space Telescope Science Institute, Baltimore, MD}   
\email{meixner@stsci.edu}
\author{J. S. Gallagher, III} 
\affil{University of Wisconsin, Madison, WI} 
\email{jsg@astro.wisc.edu}
\author{M. S. Oey}
\affil{Department of Astronomy, University of Michigan, Ann Arbor, MI} 
\email{msoey@umich.edu}
\author{A. Pasquali} 
\affil{Max-Planck-Institut f$\ddot{u}$r Astronomie, Heidelberg, Germany} 
\email{pasquali@mpia-hd.mpg.de}
\author{L. J. Smith\altaffilmark{2}} 
\affil{University College London, London, UK} 
\email{lsmith@stsci.edu}
\author{M. Tosi} 
\affil{INAF-Osservatorio di Bologna, Bologna, Italy} 
\email{monica.tosi@oabo.inaf.it}
\author{R. Walterbos} 
\affil{New Mexico State University, Las Cruces, NM}
\email{rwalterb@nmsu.edu}
\altaffiltext{1}{Space Telescope Operation Division, ESA, Baltimore, MD}
\altaffiltext{2}{Space Telescope Science Institute, Baltimore, MD}

\begin{abstract}
NGC 602 is a young stellar cluster located in a peripheral region of the Small Magellanic Cloud known as the wing.  Far from the main body of the galaxy and abutting the Magellanic Bridge, the SMC's wing is characterized by low gas and stellar content.  With deep optical imaging from the Advanced Camera for Surveys (ACS) aboard the Hubble Space Telescope (HST), we have discovered an extensive pre-Main Sequence (PMS) population, with  stellar masses in the range 0.6--3~$M_{\sun}$.  These low mass PMS stars formed coevally with the central cluster about 4~Myr ago.  Spitzer Space Telescope (Spitzer) images of the same region from the Infrared Array Camera (IRAC) also reveal a population of Young Stellar Objects (YSOs), some of which are still embedded in nebular material and most of which likely formed even more recently than the young stars detected with HST/ACS imaging.  We infer that star formation started in this region $\sim$4~Myr ago with the formation of the central cluster and gradually propagated towards the outskirts where star formation is presently ongoing.
\end{abstract}


\section{Introduction}

We discuss observations of the young star cluster NGC~602 associated with the highly structured HII region N90 \citep{hen55}.  
NGC~602 is located in a region of low stellar density in the periphery of the Small Magellanic Cloud (SMC) in an area known as the ``wing'' of the SMC, which stretches between Shapley's wing \citep{sha40} and the main body of the galaxy.  This region contains a population of hot and young stars, aged 10--60 Myr \citep{wes71, dem98, cou95, mas00}, and it is believed to display the metallicity characteristics of the SMC: Z $\sim$ 0.004 \citep{lee05, rol99}.

Among the characteristic properties of the SMC, its low metallicity and low dust content \citep[1/30 that of the Milky Way;][]{sta00} impose intriguing boundary conditions on our current understanding of star formation. With the objective of characterizing star formation in this peculiar environment, we  have carried out an extensive Hubble Space Telescope (HST) and Spitzer Space Telescope \citep[Spitzer;][]{wer04} investigation  of the stellar content and star formation acivity in a number of young clusters.  Here, we present our first results for NGC~602.

\section{Observations}

Observations of NGC~602 (RA~= 1:29:31, DEC~= $-$73:33:15, J2000) were taken in the optical and mid-infrared, with  the Advanced Camera for Surveys (ACS) on the HST (Proposal 10248, also including NGC~346) and the Infrared Array Camera (IRAC) on Spitzer (Program ID 125). 

In the optical, the ACS Wide Field Channel (WFC) was used for observations in three filters: F555W, F814W (approximately the Johnson~V and I~bands), and F658N (H$\alpha$).  In~Plate \ref{acsplate}, we show the fully reduced and multidrizzled F555W, F814W, F658N color composite image of NGC 602.  The image covers a region of  $200'' \times 200''$, corresponding to a linear size of 58~pc $\times$ 58~pc at the distance of the SMC \citep[60.6~kpc;][]{hil05}. The total integration times are 2150s for the F555W image, 2265s for the F814W image, and 1908s for the F658N image, respectively.  

All images  were reduced using the standard STScI ACS pipeline CALACS.  Photometry was performed on the drizzled images by aperture and Point Spread Function (PSF) fitting, using the IRAF DAOPHOT package.  Stars were automatically detected with DAOFIND in all frames, with the detection threshold set at 4$\sigma$ above the background for the F555W image and at 3.5$\sigma$ for the F814W image.  Spurious features, such as noise spikes, were rejected. A spatially variable PSF was computed for both the F555W and F814W images, using 110 isolated stars in different positions on the images. The correction for Charge Transfer Efficiency (CTE) was applied. The final photometry was calibrated on the ACS Vega-magnitude system, normalized to a standard 
10~pixel (0\farcs5) aperture radius, with zero points adopted from \citet{sir05}.

\begin{figure}
\plotone{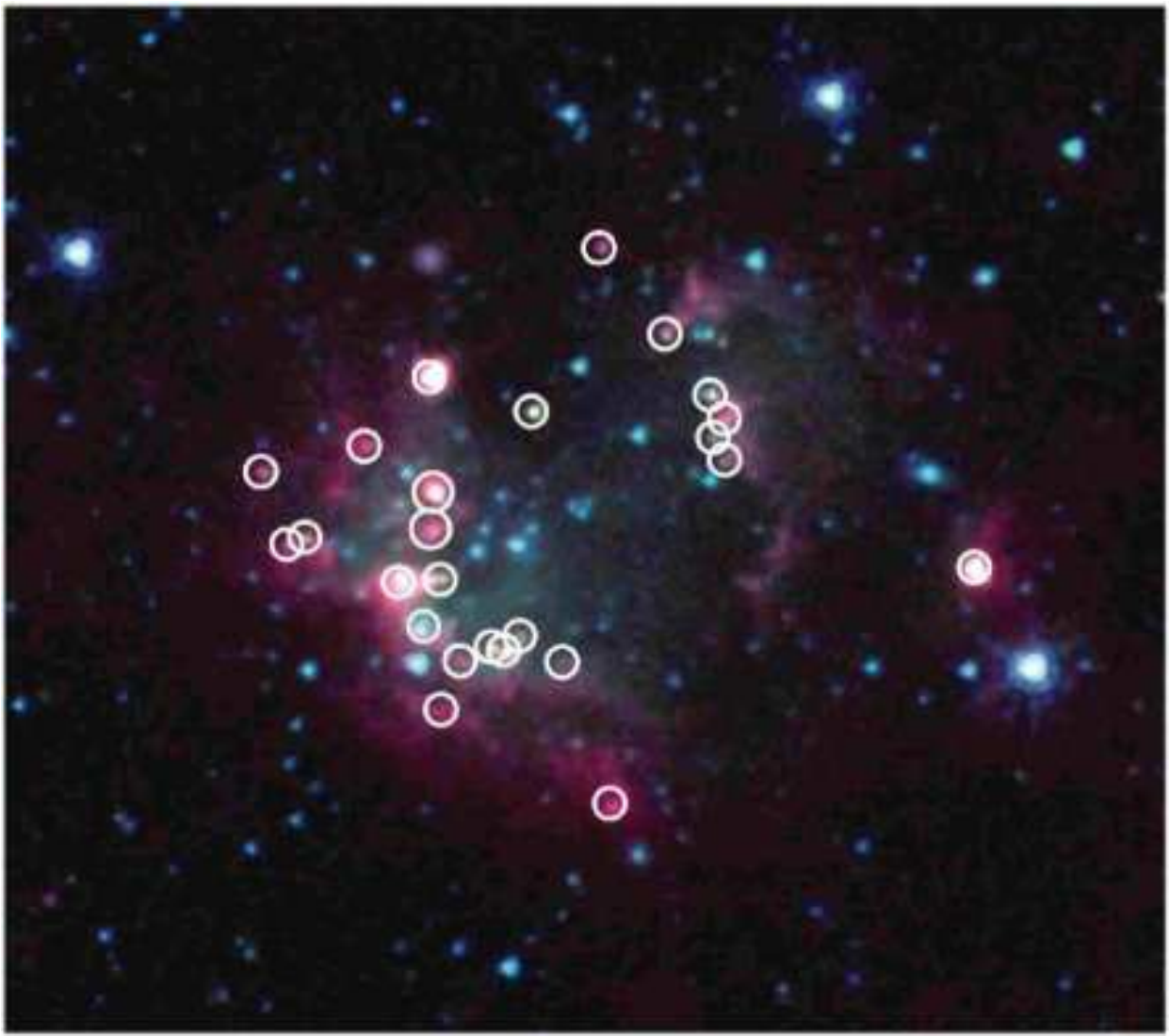}
\caption{3.6, 4.5, and 8.0 $\mu$m color composite image of the NGC~602 region, obtained with Spitzer/IRAC. White circles indicate the positions of candidate Young Stellar Objects, which appear primarily along IR-bright dust ridges and in elephant trunk formations.  The field of view is $5'\times5'$ (88~pc $\times$ 88~pc). North is up, East to the left. \label{iracplate}}
\end{figure}

In the infrared, we present images from Spitzer/IRAC in the four bands: 3.6, 4.5, 5.8, and 8.0~$\mu$m \citep{faz04}.  The final three band ([3.6], [4.5], [8.0]) composite IRAC image is shown in Plate~\ref{iracplate}.  The data were taken in 30s High Dynamic Range mode, using five dithers in the medium cycling pattern.  Each frame has an exposure time of 26.8s, for a total integration time of 134s in each band. The total field of view is approximately $5'\times5'$, corresponding to a linear size of 88~pc $\times$ 88~pc.  The S11.4.0 version of the Basic Calibrated Data (BCD) were used for constructing the IRAC mosaics, using the Spitzer Mosaicker software in each channel, along with the IRACproc package \citep{sch06}.  The overlap correction module was used to minimize instrumental offsets between frames, and the mosaics were constructed on a pixel scale of 0\farcs4/pixel. The sources were located and photometry extracted from the mosaic using the IRAF tasks DAOFIND and PHOT, using a cutoff of 4$\sigma$ to find the sources.  A radius of 1\farcs6 (4~mosaic pixels) was used in the PHOT task, and zero point corrections were applied to normalize the photometry to the nominal 10~pixel ($12''$) radius. The Vega magnitude zero points used were 24.804, 24.116, 22.058, and 22.478 for channels 1--4, respectively. 

\section{The NGC 602 Region}

The newly acquired HST/ACS and Spitzer/IRAC images offer, for the first time, a high resolution view of the NGC~602 field.  The SMC wing is a low density region and very transparent, allowing us to view background galaxies with only a red tinge from its meager dust allotment (see Plate~\ref{acsplate}).  This is the case even when looking through the center of the nebula, indicating that the central column density is insufficient for any extreme reddening in this region.

The morphology of the NGC~602 region is reminiscent of a partial ring.  Two ridges of dust and gaseous filaments outline the nebular shape towards the SE and to the NW and are highlighted by magnificent elephant trunks.  The primary stellar cluster shines in the middle of the broken ring, about 18~pc from the northwest ridge and about 9~pc from the southeast ridge.  Radiation from this concentration of the brightest cluster stars appears to have swept out from the center, creating the inner cavity of the bubble.  Many faint stars are visible on, or in close proximity to, the two ridges, indicating that star formation might still  be active there.  The ring morphology is also well defined in the Spitzer/IRAC composite image (Plate~\ref{iracplate}), where the two ridges are the brightest diffuse features detected and contain most of the brightest IR sources identified, correspondint to so-called elephant trunks and dusty pillars. 

\section{The Optical and IR Color-Magnitude Diagrams}

\begin{figure}
\plotone{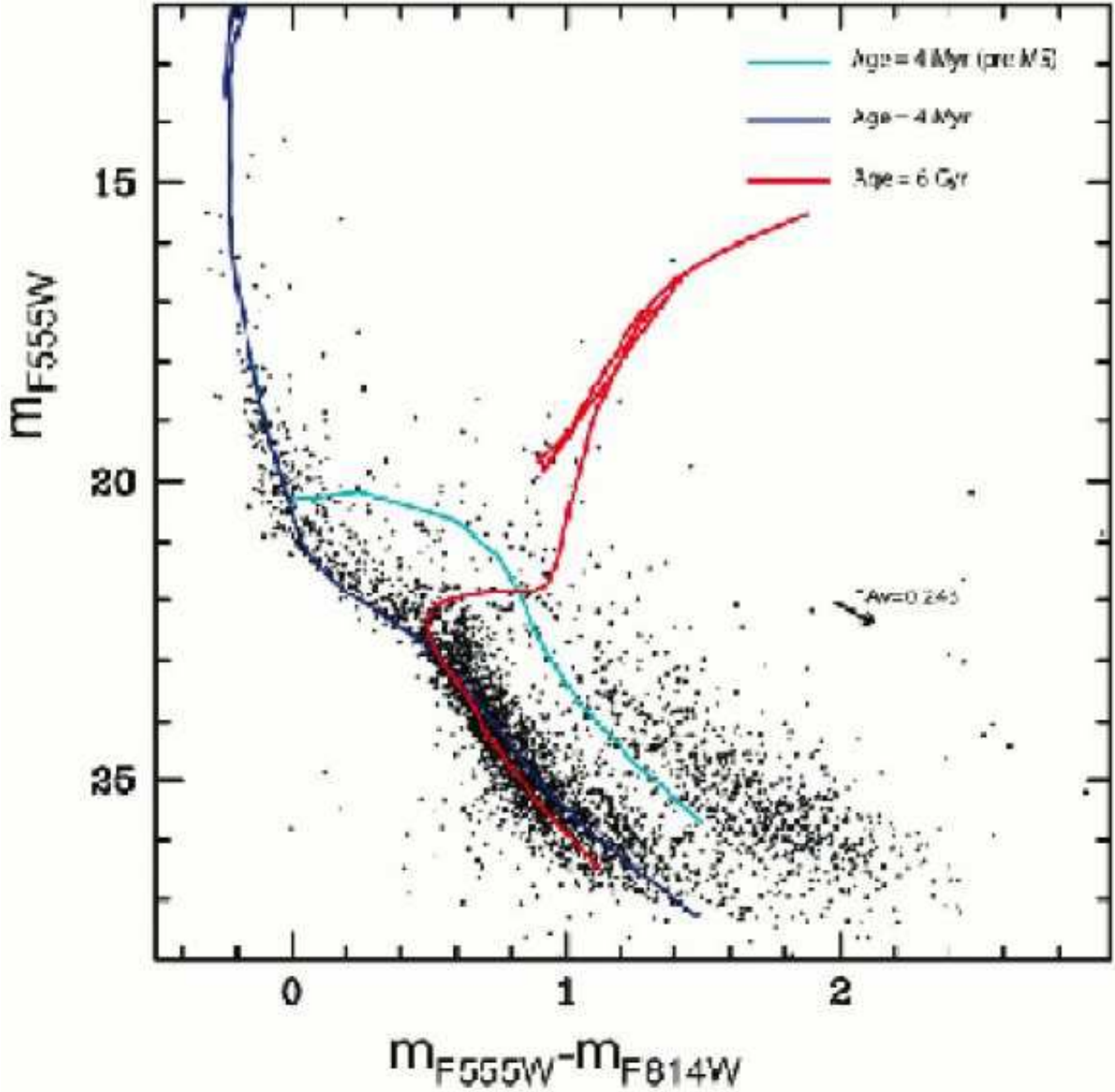}
\caption{(F555W, F814W) Color Magnitude Diagram constructed using  HST/ACS data.  Isochrones have been overlaid to determine the ages of the  different stellar populations (Blue~= 4~Myr with Z~= 0.004, Red~= 6~Gyr with Z~= 0.001).  An extinction value of E(B--V)~= 0.08 (A$_{V}= 0.248$) has been applied to isochrones.  MS and RGB isochrones from \citet{ber94}.  PMS isochrone from \citet{sie00}. \label{acscmd}} 
\end{figure}

We use Color-Magnitude Diagrams (CMDs) as a primary method of identifying the stellar populations present in the region.  The (F555W, F555W--F814W) CMD of the region is constructed using photometric data with the DAOPHOT short exposure photometric error $<$0.03 and long exposure PSF sharpness $< \mid {0.3} \mid$, thus choosing only point-like features and eliminating from the source list detections of galaxies and fuzzy brightness variations in diffuse emission regions.  This optical CMD (Fig.~\ref{acscmd}) shows a very well delineated Main Sequence (MS) down to V~= 26.5.  The upper part of the MS indicates the presence of a very young population. The lowest part of the MS has the characteristics of an older population, most likely the background field of the SMC wing.  The MS Turnoff is outlined, but not very well populated, and there is an indication of a Red Giant Branch and of a Red Clump at V~= 19.5, V--I~= 1.  In addition, we detect a sizable population of faint and red stars located to the right of the MS (V $>$ 22, 1 $<$ V--I $<$ 2).  The magnitude and color of this group of stars is consistent with the nature of low mass  (0.6--3~$M_{\sun}$) PMS stars.

Isochrone fitting allows us to estimate the ages of these different stellar populations. As can be seen in Fig.~\ref{acscmd}, the bright blue part of the cluster MS is well fit by a \citet{ber94} isochrone of 4~Myr, Z~= 0.004. The fainter MS belongs mostly to an older population, dated at $\sim$6~Gyr, with Z~= 0.001, consistent with the SMC field on the edge of the wing near the Magellanic Bridge \citep{rol99,leh01} as determined from complementary HST/ACS observations of a nearby ``featureless'' region.  Isochrone fitting also allows us to conclude that the rich PMS population \citep{sie00} is coeval with the cluster Main Sequence with an age of about 
4~Myr, and these individual low-mass objects have not yet reached the MS.  In all cases, we have applied a low E(B--V)~= 0.08 
(A$_{V} = 0.248$) reddening factor to the isochrones, as determined from ground-based observations in B, V, and~I.

\begin{figure}
\plotone{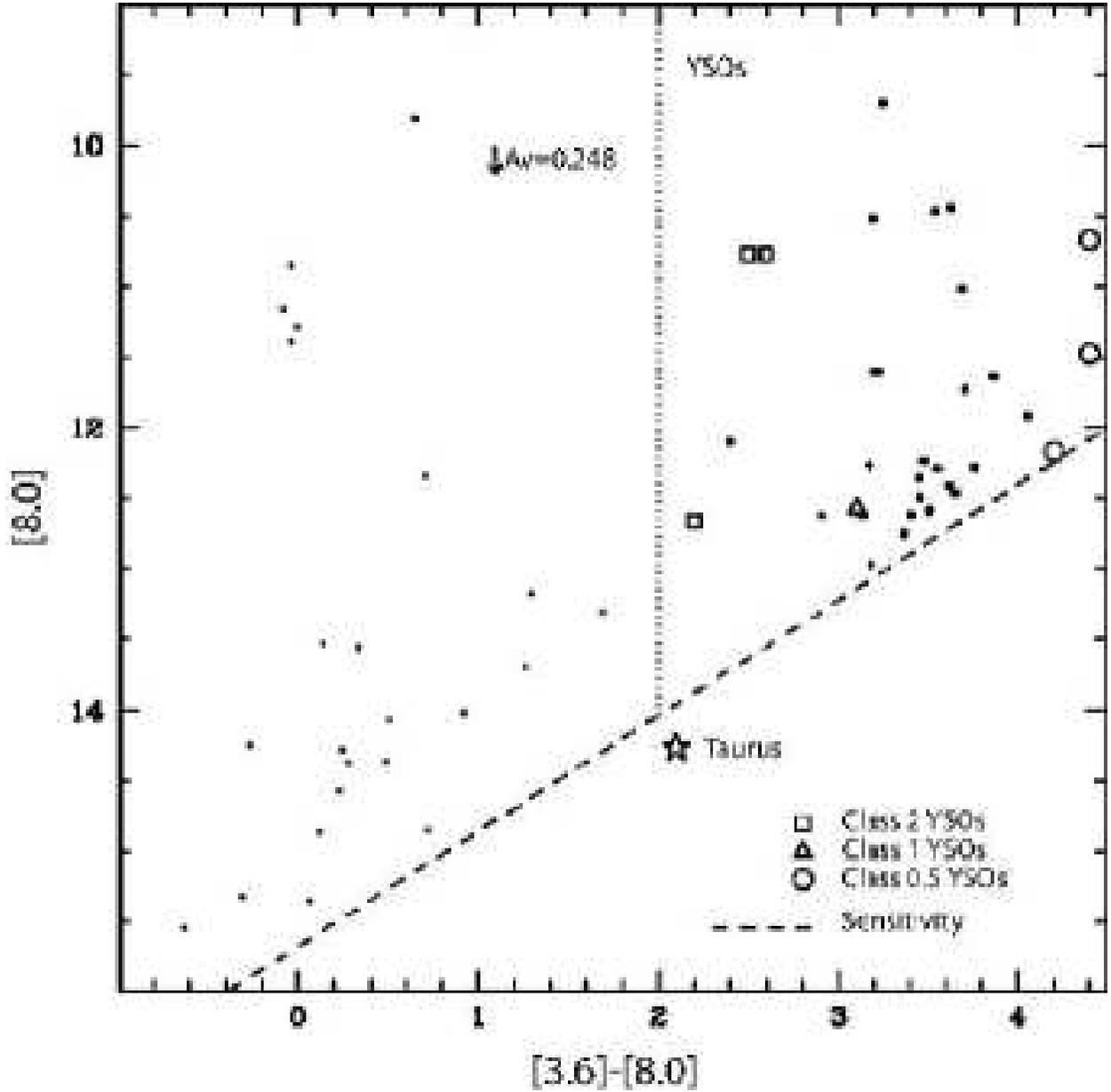}
\caption{([8.0], [3.6]--[8.0]) IRAC Color-Magnitude Diagram. The dot-dash line encloses the region of color-magnitude space in which we expect to find Young Stellar Objects.  YSO models are for stellar masses greater than about 6~$M_{\sun}$ and include multiple viewing angles.  Sources with ([3.6]--[8.0]) $<$ 2 are primarily MS stars.  Taurus magnitude adapted from \citet{mei06}. \label{iraccmd}}
\end{figure}

The IRAC data allow us to sample a population of young stellar objects (YSOs) even younger than the PMS population we detect in the optical images.  After examining five CMDs, we identify 25 good YSO candidates.  Objects readily identifiable as background galaxies in the optical images are manually removed from the object list, leaving 51 point-like sources with measurable flux in the three bands [3.6], [5.8], and [8.0].  We use YSO models for a range of ages, masses, and viewing angles \citep{whi03,whi04} to help with the identification of the color-magnitude space in which we expect to find YSOs.  The ([8.0], [3.6]--[8.0]) IRAC CMD is representative and is shown in Fig.~\ref{iraccmd}.

Young Stellar Object (YSO) models span an evolutionary range from Class 0 ($\sim$10$^4$~yr) to Class~II ($10^6$--$10^7$~yr). Class~0 YSOs are in the early protostellar collapse phase, with T~$\sim$ 30K. Class~I sources are at a later stage of protostellar collapse which lasts $\sim$10$^5$~yr.  Class~II sources are characterized by the presence of  excess infrared emission above a stellar photosphere \citep{whi03, whi04}. Our YSO candidates fall in the same Color-Magnitude space as Class~0.5 and Class~I YSO models, indicating that we are finding recent, possibly still ongoing, star formation which began about 1~Myr ago.  The low-mass PMS population that we see in the HST/ACS data corresponds to more evolved, likely Class~II  objects. However, these objects have very low masses ($0.6$ -- $3~M_{\sun}$) and magnitudes which fall below our Spitzer/IRAC detection limit.  

\section{The Spatial Distribution of Young Stars}

\begin{figure}
\plotone{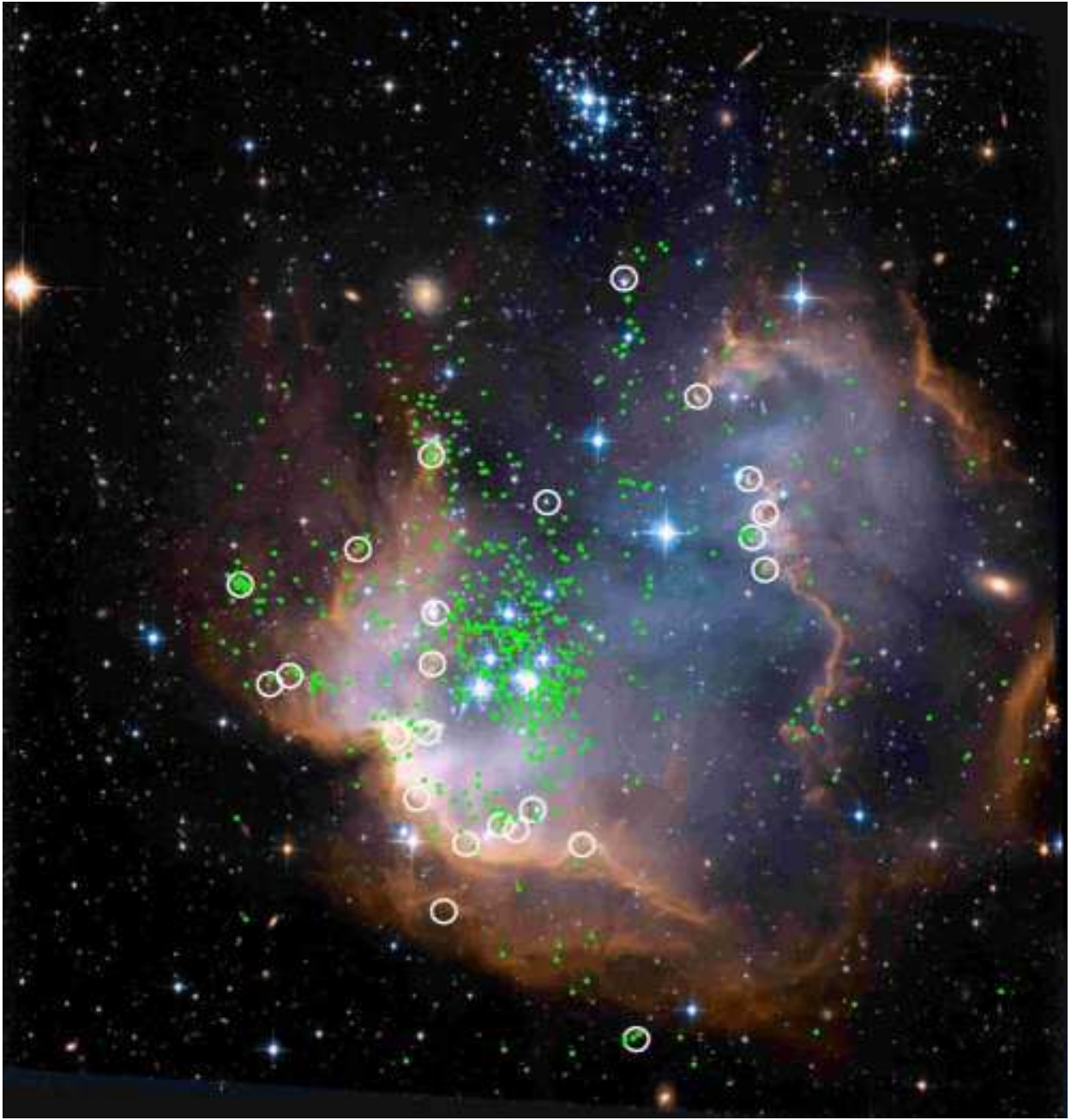}
\caption{Spatial distribution of PMS stars (small green dots) and candidate YSOs (white circles) marked on the 3-color HST/ACS image.  Note that the PMS stars are concentrated near cluster center, while the YSO candidates coincide with dusty ridges and pillar-like structures. \label{marked}}
\end{figure}

Fig.~\ref{marked} shows the spatial distribution of PMS stars (green dots) as detected in the optical images and YSOs (white circles; also shown on Plate~\ref{iracplate}) selected from the IR dataset. The pre-Main Sequence population is concentrated in the center of the cluster, where the brightest and most massive stars, those with spectral types~O and~B \citep{wes64,hut91}, are also found. To the East, to the North, and along the SE ridge, we see a number of compact knots of faint red stars in regions veiled by dust.  Outside the central cluster, PMS stars appear generally concentrated in compact, structured groups, similar to the sub-clusters detected in the young SMC region NGC~346 \citep{not06,sab07}.

Farther from the cluster center, outside the main nebular structure, the stellar population becomes dominated by the SMC field, and the spatial density of PMS stars drops considerably.  For example, the bright sub-cluster located due north of  the central cluster, near the top edge of the field of view, displays a population composed exclusively of MS and Red Giant Branch (RGB) stars.

The YSOs, as detected in the IRAC observations, have a different spatial distribution than the PMS population.  Overall, the YSOs are farther from the cluster center.  This agrees with expectations, since the winds and radiation from the most massive stars appear to have effectively removed dust and gas from this inner region.  Instead, YSOs mostly appear concentrated along the NW and SE dusty ridges,  indicating that star formation is more recent and is likely still active and progressing within these two structures, forming dusty pilars remeniscent of those seen in other known star-forming regions.

\section{Conclusion: Star Formation in NGC 602}

We have shown that the combination of recently acquired deep optical HST/ACS images and IR Spitzer/IRAC images of NGC~602 suggests an intriguing chronological sequence  of star forming events in this young cluster.  The high resolution optical data show a rich population of low mass (0.6--3~$M_{\sun}$) PMS stars that formed coevally with the central cluster about 4~Myr ago.  Conversely, the brightest sources detected in the IR images have magnitudes and colors consistent with state-of-the-art model YSOs of 
Class~0.5 and Class~I.  These objects have typical lifetimes of $10^4$--$10^5$~yr, indicating that the region is likely experiencing ongoing episodes of star formation. 

The youngest objects are far from the central cluster and are concentrated along the two dusty ridges. The crossing time from the cluster center to the these ridges is between~1 and 2~Myr, and it is very plausible that the most recent star formation is triggered by instabilities caused by the earlier formation of the central cluster. It is also likely that star formation is actively continuing in the region of the ridges, where gas and dust are still abundant. 

Further work is necessary to fully characterize the star formation in this region.  This includes spectroscopic characterization of the PMS stars and YSOs, classification of individual YSO's by fitting spectral energy distributions to models, and spectroscopic observations of the gas kinematics.

\begin{figure}
\plotone{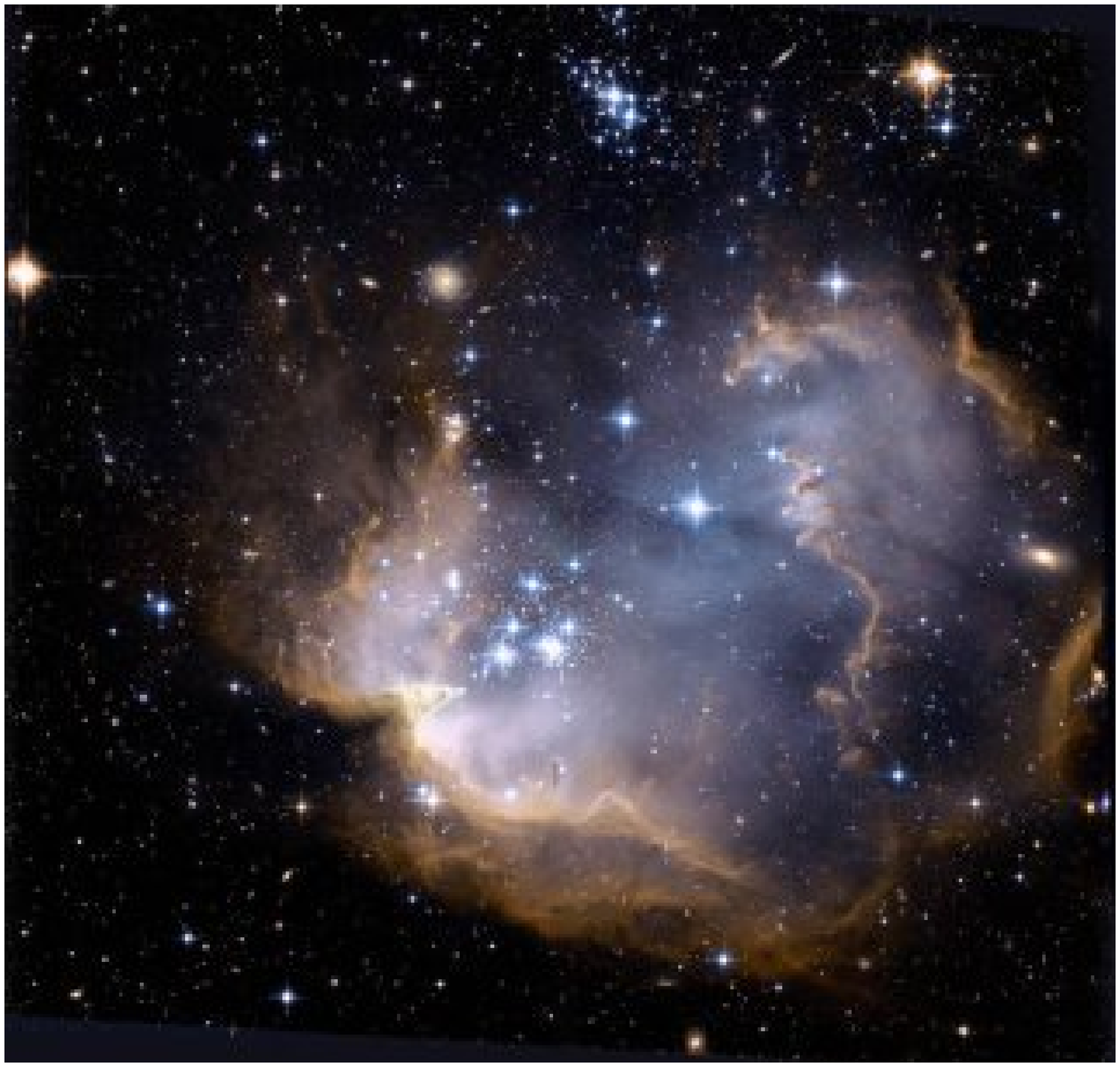}
\caption{F555W, F814W, F658N color composite image of the NGC~602 region, obtained with HST/ACS.  The field of view is $200''\times200''$. North is up, East to the left.  (Image production: Hubble Heritage team) \label{acsplate}}
\end{figure}

\acknowledgments
We wish to thank Lars Christensen (ESA) and Zolt Levay and Lisa Frattare (STScI) for producing the beautiful color composite image of NGC~602 presented in Plate~\ref{acsplate}.  HST funding came from STScI GO grant GO-10248.07-A.  This work is also based in part on observations (Program 125, Request Key 12485120) made with the Spitzer Space Telescope, which is operated by the Jet Propulsion Laboratory, California Institute of Technology, under NASA contract 1407.  Support for this work was provided by NASA through contract 1256790 issued by JPL/Caltech. 
{\it Facilities:} \facility{HST (ACS)}, \facility{Spitzer (IRAC)}.

\clearpage

\clearpage

\clearpage

\clearpage

\clearpage

\begin{thebibliography}{}
\bibitem[Bertelli et~al.(1994)]{ber94} Bertelli, G., Bressan, A., Chiosi, C., Fagotto, F., \& Nasi, E. 1994, A\&AS, 106, 275
\bibitem[Courtes et~al.(1995)]{cou95} Courtes, G., Viton, M., Bowyer, S., Lampton, M., Sasseen, T.~P., \& Wu, X.-Y. 1995, A\&A, 297, 338
\bibitem[Demers \& Battinelli(1998)]{dem98} Demers, S., \& Battinelli, P. 1998, \aj, 115, 1472
\bibitem[Fazio et~al.(2004)]{faz04} Fazio, G., et al. 2004, \apjs, 154, 10 
\bibitem[Henize(1955)]{hen55} Henize, K. 1955, \apjs, 2, 315
\bibitem[Hilditch, Howarth, \& Harries(2005)] {hil05} Hilditch, R. W., Howarth, I. D., \& Harries, T. J. 2005, \mnras, 357, 304
\bibitem[Hutchings(1991)]{hut91} Hutchings, J. B., Thompson, I. B., Cartledge, S., \& Pazder, J. 1991, \aj, 101, 933
\bibitem[Lee et~al.(2005)]{lee05} Lee, H., et al. 2005, AAS, 207, 113.11
\bibitem[Lehner et~al.(2001)]{leh01} Lehner, N., Sembach, K. R., Dufton, P. L., Rolleston, W. R. J., \& Keenan, F. P. 2001, \apj, 551, 781
\bibitem[Massey, Waterhouse, \& DeGioia-Eastwood(2000)]{mas00} Massey, P., Waterhouse, E., \& DeGioia-Eastwood, K. 2000, \aj, 119, 2214
\bibitem[Meixner et~al.(2006)]{mei06} Meixner, M., et al. 2006, \aj, 132, 2268
\bibitem[Nota et~al.(2006)]{not06} Nota, A., et al. 2006, \apj, 640L, 29
\bibitem[Rolleston et~al.(1999)]{rol99} Rolleston, W. R. J., Dufton, P. L., McErlean, N. D., \& Venn, K. A. 1999, A\&A, 348, 728
\bibitem[Sabbi et~al.(2007)]{sab07} Sabbi, E., et al. 2007, \aj, 133,44
\bibitem[Schuster et~al.(2006)]{sch06} Schuster, et al. 2006, SPIE Proc. 6270, 65
\bibitem[Siess, Dufour, \& Forestini(2000)]{sie00} Siess, L., Dufour, E., \& Forestini, M. 2000, A\&A, 358, 593
\bibitem[Sirianni et~al.(2005)]{sir05} Sirianni, M., et al. 2005, \pasp, 117, 1049
\bibitem[Shapley(1940)]{sha40} Shapley, H. 1940, BHarO, 914, 8
\bibitem[Stanimirovic et~al.(2000)]{sta00} Stanimirovic, S., Staveley-Smith, L., van der Hulst, J. M., Bontekoe, T. J. R, Kester, D. J. M., \& Jones, P. A. 2000, \mnras, 315, 719
\bibitem[Werner et~al.(2004)]{wer04} Werner, M. W., et al. 2004, \apjs, 154, 1
\bibitem[Westerlund(1964)]{wes64} Westerlund, B. E. 1964, \mnras, 127, 429
\bibitem[Westerlund \& Glaspey(1971)]{wes71} Westerlund, B. E., \& Glaspey, J. 1971, A\&A, 10, 1
\bibitem[Whitney et~al.(2003)]{whi03} Whitney, et al. 2003, \apj, 598, 1079
\bibitem[Whitney et~al.(2004)]{whi04} Whitney, et al. 2004, \apjs, 158, 315
\end{thebibliography}
\end{document}